\documentclass[pre,twocolumn,showpacs]{revtex4}

\newcommand{\g}[1]{{\bf {#1}}}

\begin{document}

\title{Current-sheet formation in incompressible electron magnetohydrodynamics}
\author{V.P. Ruban}
\email{ruban@itp.ac.ru}
\affiliation{L.D.Landau Institute for Theoretical Physics,
2 Kosygin Street, 117334 Moscow, Russia}
\affiliation{Optics and Fluid Dynamics Department, 
Ris\o ~National Laboratory, DK-4000 Roskilde, Denmark}
\date{\today}
\begin{abstract}
The nonlinear dynamics of axisymmetric, as well as helical,
frozen-in vortex structures is investigated by the Hamiltonian method 
in the framework of ideal incompressible electron magnetohydrodynamics. 
For description of current-sheet 
formation from a smooth initial magnetic field, local and nonlocal 
nonlinear approximations are introduced and partially analyzed that are 
generalizations of the previously known exactly solvable local model neglecting 
electron inertia. Finally, estimations are made that predict finite-time
singularity formation for a class of hydrodynamic models 
intermediate between that local model and the Eulerian hydrodynamics.
\end{abstract}
\pacs{52.30.Cv, 52.30.Ex}
\maketitle

{\it Introduction. ---}
It is a well known fact that current sheets play exclusively important role
in plasma dynamics (see, e.g., 
\cite{Biskamp1993,PP,Grauer&Marliani,Muller&Biskamp2000}
and references therein). However, analytical study of current sheets 
formation and their dissipative dynamics is a very difficult problem
in the framework of usually used nonlinear (and also nonlocal in the
incompressible limit) equations of motion of plasmas. 
That concerns the usual magnetohydrodynamics (MHD), 
the electron magnetohydrodynamics (EMHD),
as well as the multi-fluid models of plasmas. So, up today we do not have a
mathematically clear answer on the question whether the current density will
become singular in a finite time or its growth can be only exponential
in these systems. Numerical simulations remain to be the main tool for 
obtaining quantitative results  
\cite{Biskamp1993,PP,Grauer&Marliani,Muller&Biskamp2000,Biskamp_et_al,
Porcelli_et_al, Boffetta_et_al}. 
Therefore, an important role for theoretical understanding of current sheets 
dynamics can be played by local nonlinear approximations, that sometimes
have exact solutions describing formation of singularities.
An example of such relatively simple, approximate differential equation 
for the magnetic field $\g{B}(\g{r},t)$ is (see, e.g., \cite{KChYa1990}
for derivation and explanation)
\begin{equation}\label{theLocalModel}
\g{B}_t=-\frac{c}{4\pi e}
\mbox{curl}\left[\frac{\mbox{curl\,}\g{B}}{n}\times\g{B}\right].
\end{equation}
This equation describes the motion of magnetic structures in EMHD on length 
scales much larger than the inertial electron skin-depth, while the main part 
of the energy is concentrated in the magnetic field, with the kinetic energy 
of the electron fluid motion being much smaller.  
The equation (\ref{theLocalModel}) has been extensively 
exploited, for instance, to study fast penetration of magnetic field into 
plasmas due to the Hall effect \cite{Fruchtman_at_al,Kalda}, 
as well as rapid dissipation of magnetic fields in laboratory and 
astrophysical conditions \cite{VChO}. The interest to this equation is 
explained, in particular, by the fact that axisymmetric configurations with 
$\g{B}\parallel\g{e}_\varphi$ have been found exactly solvable 
(see \cite{KChYa1990,Fruchtman_at_al,Kalda,VChO}). 
In this geometry, the equation of motion is 
reduced to the well known one-dimensional Hopf equation, 
that should be solved independently for each value of the radial coordinate. 
The mechanism of singularity formation in these solutions is connected
simply with breaking in a finite time of the magnetic field profile. The
magnetic field itself does not become infinite, but its curl tends to the
infinity at some point of the axial cross-section. Inclusion of dissipative
terms into the equation stops the breaking, but instead of multi-valued profile,
a shock forms, the length of which increasing with time.
The shock is a cross-section of a current sheet. 

The main purpose of present work is to extend the analysis of such axisymmetric 
flows by consideration additional nonlinear effects caused by electron inertia.
They either play role of small corrections for long-scale flows, 
or, when shock becomes narrow, change drastically the dynamical behavior 
by smoothing the transport velocity field.
This situation is quite different in comparison with the self-similar 
EMHD solutions discussed in Ref. \onlinecite{BPS1992}.
Also, the flows with other geometrical symmetry are considered below in the 
approximation (\ref{theLocalModel}), when all the 
frozen-in magnetic lines have helical shapes with a same spatial period along 
$z$-direction. In this case different level contours of the axial component 
of the magnetic field rotate in a perpendicular plane with different angular 
velocities, thus producing the shock.
Finally, we use the developed technique to predict formation of
finite time singularities of the shock type
in a class of hydrodynamic systems that are in some sense
intermediate between the model (\ref{theLocalModel}) and the usual Eulerian
hydrodynamics. That is a contribution towards general understanding 
of possible mechanisms for singularities in hydrodynamics, that is a
long-standing theoretical problem.

{\it Incompressible two-fluid model. ---}
Before the main consideration, it is useful to recall the place of EMHD among 
different hydrodynamical plasma models \cite{KChYa1990}. 
If there are only two kinds of particles in the plasma --- 
negatively charged electrons with the mass $m$ 
and positively charged ions with the mass $M$,
then the most general is the two-fluid model, which contains MHD, EMHD, and
Hall MHD as special cases. Let the equilibrium concentration of particles
of each sort be equal to $n$. If the temperature of the system is sufficiently
large, $nT\gg\g{B}^2$, then for slow 
vortical flows one can neglect deviations of the concentrations from $n$ 
(the quasi-neutrality condition), and believe the velocity fields  
divergence-free in homogeneous case: $(\nabla\cdot\g{v}^\pm)=0$. 

Temporary, we will not take into account dissipative processes. Thus,
application of the canonical formalism becomes possible 
\cite{ZK97, M98, KR2000PRE}, which makes the analysis more compact. 
With appropriate choice for the length scales 
($\sim d_+=(Mc^2/4\pi e^2n)^{1/2}$) and for the mass scales ($\sim M$), 
the Lagrangian functional of the incompressible two-fluid
model, in the absence of an external magnetic field, takes the form
\begin{equation}\label{L_mu_incompr}
{\cal L}_\mu\{\g{v}^+,\g{v}^-\}=\int\frac{d^3\g{k}}{(2\pi)^3}\left[
\frac{|\g{v}^+_{\g{k}}|^2}{2}+\mu\frac{|\g{v}^-_{\g{k}}|^2}{2}+
\frac{|\g{v}^+_{\g{k}}-\g{v}^-_{\g{k}}|^2}{2k^2}
\right].
\end{equation}
Here $\mu=m/M$ is the only
dimensionless parameter remaining in the system. For the electron-positron 
plasma $\mu=1$, for the hydrogen plasma $\mu\approx 1/2000\ll 1$. Below we
consider the latter case. The first two terms in the expression 
(\ref{L_mu_incompr}) give the kinetic energy of the ion and electron fluids,
while the third term is the energy of the magnetic field created by the
flows of electrically charged fluids. The conditions of incompressibility 
are assumed, $(\g{v}^{\pm}_{\g{k}}\cdot\g{k})=0$.
A possible derivation of this Lagrangian 
is via using the microscopic Lagrangian of a 
system of electrically charged point particles that can be written up to 
the second order on $v/c$, as it is explained in the famous book by Landau 
and Lifshitz \cite{LL2}:
\begin{eqnarray}
{\cal L}_{\mbox{\small micro}}=\sum_a\frac{m_a\g{v}_a^2}{2}
-\frac{1}{2}\sum_{a\not =b}\frac{e_a e_b}{|\g{r}_a-\g{r}_b|}
+\sum_a\frac{m_a\g{v}_a^4}{8c^2}\nonumber\\
+\frac{1}{4c^2}\sum_{a\not =b}\frac{e_a e_b}{|\g{r}_a-\g{r}_b|}
\{\g{v}_a\cdot\g{v}_b+(\g{v}_a\cdot\g{n}_{ab})(\g{v}_b\cdot\g{n}_{ab})\},
\end{eqnarray}
where $\g{r}_a(t)$ are the positions of the point charges $e_a$, 
$\g{v}_a(t)\equiv\dot\g{r}_a(t)$ are their velocities, 
$\g{n}_{ab}(t)$ are the unit vectors in the direction between $e_a$ and $e_b$.
Introducing macroscopic averaged characteristics --- concentrations 
$n^\pm(\g{r},t)$, hydrodynamical mean velocities $\g{v}^\pm(\g{r},t)$, 
and considering the special dynamic regime of slow vortical flows 
("soft" degrees of freedom), when deviations of the concentrations
from the equilibrium are not excited, as described in 
Ref.\onlinecite{slow_flows}, one can arrive at 
the expression (\ref{L_mu_incompr}).

It is important that the variation of the action functional 
$S=\int {\cal L}_\mu dt$, which is necessary for constituting the equations
of motion, should not be performed with respect to the variations 
$\delta\g{v}^\pm(\g{r},t)$, but with respect to the variations 
$\delta\g{x}^+(\g{a},t)$ and  $\delta\g{x}^-(\g{c},t)$, where
$\g{x}^+(\g{a},t)$ and $\g{x}^-(\g{c},t)$ are incompressible Lagrangian mappings
describing the motion of points of the ion and electron fluids,
labeled by the labels $\g{a}$ and $\g{c}$.
The corresponding mathematical technique is explained, for instance, in 
Refs. \onlinecite{R,slow_flows}. 
The equations of motion of the two-fluid
incompressible system have the following structure:
\begin{equation}\label{generalizedEuler}
\frac{\partial}{\partial t}\frac{\delta 
{\cal L}_\mu}{\delta\g{v}^{\pm}(\g{r})}
=(1-\nabla\Delta^{-1}\nabla)\left[\g{v}^{\pm}(\g{r})\times
\mbox{curl\,}\frac{\delta {\cal L}_\mu}{\delta\g{v}^{\pm}(\g{r})}\right],
\end{equation}
where the operator in the parentheses on the r.h.s. is the projector
onto the functional space of divergence-free 3D vector fields \cite{R,ZK97}.
The two vector fields $\g{p}^{\pm}(\g{r})\equiv
{\delta {\cal L}_\mu}/{\delta\g{v}^{\pm}(\g{r})}$ are the canonical momenta 
by definition. In the Fourier representation they are given by the expressions
\begin{eqnarray}
\g{p}^+_{\g{k}} &=& \frac{\delta{\cal L}_\mu}{\delta\g{v}^+_{-\g{k}}}=
\left(1+\frac{1}{k^2}\right)\g{v}^+_{\g{k}}-
\frac{\g{v}^-_{\g{k}}}{k^2},\label{moment+}\\
\g{p}^-_{\g{k}} &=& \frac{\delta{\cal L}_\mu}{\delta\g{v}^-_{-\g{k}}}=
\left(\mu+\frac{1}{k^2}\right)\g{v}^-_{\g{k}}-
\frac{\g{v}^+_{\g{k}}}{k^2}.\label{moment-}
\end{eqnarray}
Below, we will need the reversal relations for the velocities through the
momenta:
\begin{equation}\label{velocity+-}
\g{v}^+_{\g{k}} = \frac{(\mu k^2 +1)\g{p}^+_{\g{k}}+\g{p}^-_{\g{k}}}
{\mu k^2 +1 +\mu},\quad
\g{v}^-_{\g{k}} = \frac{(k^2 +1)\g{p}^-_{\g{k}}+\g{p}^+_{\g{k}}}
{\mu k^2 +1 +\mu}.
\end{equation}
It is possible to reformulate the equations (\ref{generalizedEuler}) as
equations for frozen-in vortices,
\begin{equation}\label{frozen-in}
\g{\Omega}^{\pm}_t(\g{r})= 
\mbox{curl\,}\left[\mbox{curl\,}\frac{\delta {\cal H}_\mu}
{\delta\g{\Omega}^{\pm}(\g{r})}\times\g{\Omega}^{\pm}(\g{r})
\right],
\end{equation}
where the canonical vorticity fields are defined as the curls of the canonical
momenta, $\g{\Omega}^{\pm}(\g{r},t)\equiv\mbox{curl\,}\g{p}^{\pm}(\g{r},t)$,
and also the Hamiltonian functional of the system is calculated:
\begin{eqnarray}\label{H_mu_incompr}
{\cal H}_\mu\{\g{\Omega}^+\!,\g{\Omega}^-\}\!\equiv\!\int\!
\left\{\left(\g{p}^+\!\cdot\!\g{v}^+\right)
+\left(\g{p}^-\!\cdot\!\g{v}^-\right) 
\right\}d\g{r}-{\cal L}_\mu\nonumber\\
=\!\int\!\!\frac{d^3\g{k}}{(2\pi)^3}\!
\left[
\frac{(\mu k^2\!+\!1)|\g{\Omega}^+_{\g{k}}|^2\!+
\!(k^2\!+\!1)|\g{\Omega}^-_{\g{k}}|^2\!+\!
2(\g{\Omega}^+_{\g{k}}\!\cdot\!\g{\Omega}^-_{-\g{k}})}{2k^2(\mu k^2 +1 +\mu)}
\right]\!.
\end{eqnarray}

It is clear that in the problem under consideration
there are two separated dimensionless scales 
of inverse length, $k_+\sim 1$ and $k_-\sim 1/\lambda$, where 
$\lambda=\sqrt{\mu}$ is the electron inertial skin-depth (normalized to $d_+$). 
Since $\lambda^2\ll 1$, we may write with very good accuracy
${\cal H}_\mu\{\g{\Omega}^+,\g{\Omega}^-\}\approx
{\cal H}_\lambda\{\g{\Omega}^+,\g{\Omega}^-\}$, where
\begin{widetext}
\begin{equation}\label{H_lambda}
{\cal H}_\lambda\{\g{\Omega}^+,\g{\Omega}^-\}=
\frac{1}{2}\int\frac{d^3\g{k}}{(2\pi)^3}\Big[G_{++}(k)|\g{\Omega}^+_{\g{k}}|^2
  +    G_{--}(k)|\g{\Omega}^-_{\g{k}}|^2+
     2G_{+-}(k)(\g{\Omega}^+_{\g{k}}\cdot\g{\Omega}^-_{-\g{k}})\Big],
\end{equation}   
\end{widetext}  
\begin{eqnarray}
G_{++}(k)=\frac{1}{k^2},\quad
G_{+-}(k)=\left(\frac{1}{k^2}-\frac{1}{k^2+\lambda^{-2}}\right),\nonumber\\
G_{--}(k)=\left(\frac{1}{k^2}+\frac{1}{1+\lambda^2 k^2}\right).
\end{eqnarray}

Depending on the typical spatial scale of the vortices, several dynamical regimes 
are possible in this system. The small and moderate wave number region,
$k <\sim 1$, corresponds to the Hall MHD, and in the special limit
$|\g{\Omega}^+ +\g{\Omega}^-|\ll |\g{\Omega}^+|, |\g{\Omega}^-|$, we have
here the usual MHD. The region 
$1\ll k <\sim 1/\lambda$, under the extra condition
$|\g{\Omega}^+|\ll|\g{\Omega}^-|$, corresponds to the EMHD \cite{KChYa1990}. 
For the flows with larger typical wave numbers, $k\gg 1/\lambda$, 
the magnetic effects become relatively un-significant, and the system
(\ref{H_lambda}) is broken into two weakly interacting subsystems, each of them 
being approximately described by the ordinary Eulerian hydrodynamics,
since $G_{--}(k)\approx 1/\lambda^2k^2$, 
$G_{+-}(k)\approx 1/\lambda^2k^4\ll G_{++}(k),G_{--}(k)$ in this region.

{\it Axisymmetric large-scale EMHD flows. ---}
Let us now consider the subset of solutions, for which the ion canonical 
vorticity is identically equal to zero, $\g{\Omega}^+=0$, 
and the electron vorticity $\g{\Omega}^-_{\g{k}}$
is concentrated in the range $1\ll k \ll 1/\lambda$ of the wave numbers, where
the Green's function $G_{--}(k)$ is almost flat: $G_{--}(k)\approx 1 $.
Practically this corresponds to the condition $3<\sim k <\sim 20$.
For EMHD model this is the long-scale region, where $\g{\Omega}^-$ is 
proportional to the magnetic field in the leading order. 
It should be emphasized that with $\g{\Omega}^+=0$ the velocity 
$\g{v}^+$ of the ion component is not exactly zero, however, it is much smaller
than the velocity $\g{v}^-$ of the electron component, as it becomes clear from
consideration of the Eqs. (\ref{velocity+-}) with 
$\g{p}^+=0$. In the main approximation, the Hamiltonian for the electron
vorticity takes the very simple form
\begin{equation}\label{H_local_0}
{\cal H}_\lambda\{\g{0},\g{\Omega}^-\}\approx
\frac{1}{2}\int |\g{\Omega}^-|^2d\g{r},
\end{equation}
in accordance with the fact that the energy of the system is concentrated 
mostly in the magnetic field. The corresponding equation of motion is local 
and essentially coincides with Eq. (\ref{theLocalModel}):
\begin{equation}\label{dyn_local}
\g{\Omega}^-_t=\mbox{curl}\left[\mbox{curl\,}\g{\Omega}^-\times\g{\Omega}^-\right].
\end{equation}
One of remarkable properties of the equation (\ref{dyn_local}) is that 
in the case of axisymmetric flows, when
\begin{equation}\label{axisymmetric}
\g{\Omega}^-(\g{r},t)=\omega^-(q,z,t)[\g{e}_z\times\g{r}],
\end{equation}
where $q=(x^2+y^2)/2$, we have the exactly solvable Hopf equation for the
function $\omega^-(q,z,t)$ \cite{KChYa1990}:
\begin{equation}\label{Hopf}
\omega^-_t+2\omega^- \omega^-_z =0.
\end{equation}
The solution of the equation (\ref{Hopf}) at $t>0$ is constructed from the
initial function $\omega^-_0(q,z)$ by the shift of each level contour 
$\omega^-_0(q,z)=w$ along $z$-axis on the value $2wt$, that makes possible
breaking of the profile after some time. 
Not long before the moment of the singularity formation,  
the equation (\ref{dyn_local}) becomes non-applicable. For correction, it
is sometimes sufficient  
to add into the r.h.s. of the equation (\ref{dyn_local}) the only
linear dissipative term $(e^2n/M\sigma)\Delta \g{\Omega}^-$, which takes into
account a finite electrical conductivity $\sigma$ \cite{KChYa1990}.
In this case the equation for the function $\omega^-(q,z,t)$ looks as follows:
\begin{equation}\label{Burgers}
\omega^-_t+2\omega^- \omega^-_z =\frac{e^2n}{M\sigma}\left(
2q\omega^-_{qq}+4\omega^-_q+\omega^-_{zz}\right).
\end{equation}
In order to justify the neglect by dispersive effects, the typical values of 
$\omega^-$ should not be too large:
$\omega^-<\sim{e^2n}/{2\lambda M\sigma}\approx {10 e^2n}/{M\sigma}$.
With this condition the width of the current sheet will remain several times
larger than the dispersive length $\lambda$. Otherwise, it is necessary to take
into account subsequent terms in the expansion of the Green's function $G_{--}(k)$ 
on powers of $\lambda^2 k^2$ (we may neglect the term $1/k^2$ as previously,
since $k\gg 1$):
\begin{eqnarray}
G_{--}(k)&\approx& 1-\lambda^2k^2+(\lambda^2k^2)^2+\dots,\label{G--expansion}\\
{\cal H}_\lambda\{\g{0},\g{\Omega}^-\}&\approx&
\frac{1}{2}\int \g{\Omega}^-\cdot\left(
1+\lambda^2\Delta+\dots\right)\g{\Omega}^-d\g{r}.
\end{eqnarray}

Let us consider the axisymmetric flows like (\ref{axisymmetric}). 
It is useful to note that in the absence of dissipation, as follows from   
Eqs. (\ref{frozen-in}), the dynamics of the functions $\omega^\pm(q,z,t)$
possesses the remarkable structure:
\begin{equation}\label{2Daxisymm}
\omega^\pm_t+
\left({\delta{\cal H}_*}/{\delta\omega^\pm}\right)_q\omega^\pm_z-
\left({\delta{\cal H}_*}/{\delta\omega^\pm}\right)_z\omega^\pm_q=0,
\end{equation}
where ${\cal H}_*\{\omega^+\!,\omega^-\}\!=\!(1/2\pi)
{\cal H}_\mu\{\omega^+[\g{e}_z\!\times\!\g{r}],
\omega^-[\g{e}_z\!\times\!\g{r}]\}$. 
Thus, each of the functions $\omega^\pm(q,z,t)$
is transported by its own, divergence-free in $(q,z)$-plane, 
two-dimensional velocity field, the
stream-function of which coinciding with the corresponding variational 
derivative of the Hamiltonian. The same Poisson structure governs the
ideal hydrodynamics in Cartesian plane \cite{ZK97}. 

Using the expression for the $\Delta$-operator in $(q,z)$-coordinates, 
\begin{equation}\label{Laplacian}
\Delta \{f(q,z)[\g{e}_z\times\g{r}]\}=
(2qf_{qq}+4f_q+f_{zz})[\g{e}_z\times\g{r}],
\end{equation}
we easily obtain the asymptotic expansion (for simplicity, we write $\omega$ 
instead of $\omega^-$ in the two following equations)
\begin{equation}\label{2Dexpansion}
{\cal H}_*\{0,\omega\}=\int  \omega
\left[q+\lambda^2(2\partial_q q^2\partial_q+q\partial_z^2)+\dots\right]
\omega\, dq\,dz
\end{equation}
and the corresponding conservative equation of motion
\begin{eqnarray}\label{corrections}
\omega_t+2\omega\omega_z+2\lambda^2
\Big[-(2q^2\omega_{qqz}+4q\omega_{qz}+q\omega_{zzz})\omega_q\nonumber\\ 
+(8q\omega_{qq}+4\omega_q+\omega_{zz}
+2q^2\omega_{qqq}+q\omega_{zzq})\omega_z
\Big]=0,
\label{Hopf++}
\end{eqnarray}
where the nonlinear dispersive terms are explicitly written in the first order
on $\lambda^2$. The dissipation can be taken into account like in the r.h.s. 
of the Eq.(\ref{Burgers}).

In the special case when $\omega^-$ is only slowly dependent on the radial 
coordinate $q$, but strongly depends on the axial coordinate $z$, 
the expansion of $G_{--}(k)$ on the powers of $\lambda^2(k_x^2+k_y^2)$
is appropriate:
\begin{equation}\label{slow_q_dependence}
G_{--}(k)\approx\frac{1}{1+\lambda^2k_z^2}
-\frac{\lambda^2(k_x^2+k_y^2)}{(1+\lambda^2k_z^2)^2}+\dots
\end{equation}
Then in the leading order the equation of motion for $\omega^-(z,t)$ 
becomes nonlocal integral-differential:
\begin{equation}\label{smoothing}
\omega^-_t(z,t)+\omega^-_z(z,t)\lambda^{-1}\int_{-\infty}^{+\infty}
\omega^-(\xi,t)e^{-|z-\xi|/\lambda}d\xi=0.
\end{equation}
For long-scale profiles of $\omega^-$ this equation is approximately reproduced 
by Eq. (\ref{Hopf}), but in addition it is able to describe changing of 
the steeping regime from explosive $|\omega^-_z|_{\max}\sim (t_*-t)^{-1}$ 
to exponential $|\omega^-_z|_{\max}\sim \exp C(t-t_*)$ after the width 
of the shock becomes smaller than $\lambda$. The exponential growth 
of the maximum of $|\omega^-_z|$ takes place on the final stage of shock 
evolution (without dissipation) since the integral operator 
in Eq.(\ref{smoothing}) makes the transport velocity for $\omega^-$ smooth 
enough even for a very narrow shock.

{\it Shocks in helical EMHD flows. ---}
Analogously, the helical flows can be investigated, with
\begin{eqnarray}
(\Omega^-)^z=\Omega(x\cos Kz+y\sin Kz,y\cos Kz-x\sin Kz,t),\label{spiralZ}
\\
(\Omega^-)^x=-Ky(\Omega^-)^z,\qquad (\Omega^-)^y=Kx(\Omega^-)^z,\label{spiralXY}
\end{eqnarray}
that are space-periodic along $z$-direction with the period $L^z=2\pi/K$.
The general solution of Eq. (\ref{dyn_local}) for this case can also be 
obtained, since the equation of motion for the function $\Omega(u,v,t)$ is
\begin{equation} \label{rotation}
\Omega_t+2K^2\Omega(v\Omega_u-u\Omega_v)=0.
\end{equation}
This equation follows from the Hamiltonian
\begin{equation}\label{spiralHamiltonian}
{\cal H}_s\{0,\Omega\}=\frac{1}{2} \int \Omega
\left[1+K^2(u^2+v^2)+\dots\right]\Omega\, du\,dv.
\end{equation}
Each level contour $\Omega(u,v)=W$ rotates with the individual angular velocity 
$d\theta/dt=-2K^2W$, that is the reason for shock producing. 
Higher-order corrections to Eq. (\ref{rotation}) can be derived similarly to  
Eqs. (\ref{Laplacian}-\ref{corrections}). However, in this case it is not 
possible to include the dissipation into consideration in the framework of
single-function description (\ref{spiralZ}-\ref{spiralXY}), 
since magnetic diffusivity destroys helical shapes of the magnetic lines.

{\it Shocks in Hall MHD. ---}
If we would like to escape the restriction $k\gg 1$, it is necessary to
deal with the Hall MHD, the Hamiltonian of which is
\begin{eqnarray}
{\cal H}^{HMHD}\{\g{\Omega}^+,\g{\Omega}^-\}=
\frac{1}{2}\int|\g{\Omega}^-|^2d\g{r}\nonumber\\
\label{HMHD_Hamiltonian}
+\frac{1}{2}\int(\g{\Omega}^+ +\g{\Omega}^-)(-\Delta^{-1})
(\g{\Omega}^+ +\g{\Omega}^-)d\g{r}.
\end{eqnarray}
For axisymmetric flows we have
\begin{eqnarray}\label{HMHD_Hamilt_axisymm}
{\cal H}_*^{HMHD}\{\omega^+,\omega^-\}=\int (\omega^-)^2q\,dq\,dz\nonumber\\
+\frac{1}{2}\int(\omega^+ +\omega^-)\hat G(\omega^+ +\omega^-)dq\,dz,
\end{eqnarray}
where the operator $\hat G$ is defined as follows:
\begin{widetext}
\begin{equation}\label{G_operator}
\hat G f(q,z)\equiv\frac{1}{4\pi}\int(qq_1)^{1/4}
F\left(\frac{(z-z_1)^2+2(q+q_1)}{4(qq_1)^{1/2}}\right)f(q_1,z_1)\,dq_1dz_1,
\qquad
F(A)\equiv\int_0^{2\pi}\frac{\cos\varphi \, d \varphi}{\sqrt{A-\cos\varphi}}.
\end{equation}
\end{widetext}
The equations of motion can be written in the form
\begin{eqnarray}
\omega^-_t+(2\omega^- +\Psi_q)\omega^-_z -\Psi_z\omega^-_q&=&0,\\
\omega^+_t+\Psi_q\omega^+_z -\Psi_z\omega^+_q&=&0,\\
\Psi&=&\hat G(\omega^+ +\omega^-).
\end{eqnarray}
Since the nonlocal operator $\hat G$ possesses smoothing properties, analogously to 
the usual "flat" $\Delta^{-1}$-operator, the stream-function $\Psi$ is smooth
enough even where the functions $\omega^+$ and $\omega^-$ have infinite 
gradients. Therefore, the effect of the non-locality, generally speaking, 
can not overcome the tendency
towards the breaking of the function $\omega^-$ profile, at least with
moderate typical values of $\Psi$. We can suppose that with the initial data
concentrated in the region $k\sim 1$, the breaking takes place as the general
case in the Hall MHD model. 
As concerns the transition to the limit of usual MHD, on small $k\ll 1$,
and $\omega^-\ll\Psi_q$, $|\omega^+ +\omega^-|\ll |\omega^+|,|\omega^-|$,
in this case the question about breaking 
remains subtle and needs additional investigations.

{\it Estimations for the shock singularities in different models. ---}
Finally, let us note that the equation (\ref{dyn_local}) is interesting
also from a more general theoretical viewpoint. This is an example of a
3D hydrodynamic type system, where the singularity formation explicitly takes 
place in a finite time. However, the above described mechanism for the singularity
formation can not be universal for all the hydrodynamical systems. For example, it
is known that in solutions of the Euler equation no singularity can form 
in a finite time without maximum of the vorticity growing to the infinity 
\cite{BKM1984}. Simultaneously, the field of the vorticity direction must 
loose the smoothness at the singular point \cite{CF}. 
But in the cases considered here no of these two conditions is satisfied, 
but nevertheless, the singularity develops. 
We may suppose that a type of a possible singularity in some
hydrodynamic system depends on the behavior of the corresponding Green's
function at large $k$. In Eulerian hydrodynamics $G(k)=1/k^2$, while in 
the model (\ref{dyn_local}) we have $G(k)=1$. The natural question arises: 
If the Green's function has the power-like asymptotics $G(k)\sim 1/k^\gamma$, 
with a constant $\gamma$, what is the critical value $\gamma_c$ of the 
exponent that separates systems similar to Eulerian hydrodynamics from those 
similar to the model (\ref{dyn_local}), as far as the axisymmetric flows are 
concerned? The following simple estimations give the 
answer $\gamma_c =1$. Let us consider the nonlinear transport equation like 
Eq.(\ref{smoothing}) for the conserved quantity $\omega(z,t)$,
\begin{equation}\label{smoothing_general}
\omega_t(z,t)+2\omega_z(z,t)\int_{-\infty}^{+\infty}
\tilde\omega(k,t)G(k)e^{ikz}dk/2\pi=0,
\end{equation}
where $\tilde\omega(k,t)$ is the Fourier transform of the function
$\omega(z,t)$. For simplicity, it is convenient to deal with 
antisymmetric solutions 
$\omega(-z,t)=-\omega(z,t)$ that have the shape of a smooth step, with
$\lim_{z\to\mp\infty}\omega(z,t) =\pm 1$. It is natural to monitor the growth
of the quantity $s(t)\equiv-\omega_z(0,t)$. The corresponding equation of 
motion for $s$ is the following:
\begin{equation}\label{s(t)}
\dot s(t)=-2s(t)\int_{-\infty}^{+\infty}
 ik\tilde\omega(k,t)G(k)dk/2\pi.
\end{equation}
Now we note that the function $-ik\tilde\omega(k,t)$, the Fourier transform
of $-\omega_z(z,t)$, is a smooth real-valued symmetric function with the maximum 
value $2$ at $k=0$ and with a width of distribution in $k$-space of 
order $\Delta k\sim s(t)$. Therefore the approximate relation is valid,
\begin{equation}\label{dot_s(t)approx}
\dot s(t)\sim s(t)\int_0^{s(t)}G(k)dk.
\end{equation}
From here, in the case $G(k)\sim 1/k^\gamma$ with $\gamma<1$, we easily derive
the singular asymptotic behavior
\begin{equation}\label{s(t)approx}
s\sim (t_*-t)^{1/(\gamma-1)}, 
\end{equation}
while for $\gamma>1$ the integral in Eq.(\ref{dot_s(t)approx}) converges at
large values of $s$ and therefore just exponential growth takes place, 
$s\sim\exp C(t-t_*)$.

{\it Acknowledgments. ---}
This work was supported by RFBR (grant No. 00-01-00929), by the Russian State
Program of Support of the Leading Scientific Schools (grant No. 00-15-96007), 
by the Danish Graduate School in Nonlinear Science,
and by the INTAS (grant No. 00-00292).

\end{document}